# From Digital Television to Internet?
# A general technical overview of the-
# DVB- Multimedia Home Platform Specifications

**Vita Hinze-Hoare**

**Abstract**
*This paper provides a general technical overview of the Multimedia Home Platform (MHP) specifications. MHP is a generic interface between digital applications and user machines, whether they happen to be set top boxes, digital TV sets or Multimedia Pc's. MHP extends the DVB open standards. Addressed are MHP architecture, System core, and MHP Profiles.*

## 1. Introduction

Multimedia Home Platform (MHP) extends existing Digital Video Broadcasting (DVB) standards and defines a generic interface between interactive digital applications and a range of terminals such as set top boxes, integrated digital TV sets and multimedia PC's.

*"The Multimedia Home Platform attempts to adapt existing Internet and Web standards to digital Television. The aim is to provide interactive digital content that can be viewed on set top boxes and multimedia PC's."*[1]

The following areas are addressed:

- **MHP Overview**

This addresses how the MHP will be used what it will carry and the set-top box.

- **MHP Architecture**

The three-layer architecture of resources, system software and applications will be reviewed.

- **System Core**

The MHP is based around a Java virtual machine specification from Sun Microsystems to provide generic application program interface.

- **MHP Profiles**

This is a specification to provide a consistent set of features and functions. Enhanced broadcast profile concerns one-way services. Interactive broadcast profile supports the interactive services (two- way). The Internet access profile provides a gateway to the World Wide Web.

## 2. MHP Overview

MHP is produced by the Digital Video Broadcasting Project (DVB) a European based Consortium and the standards (about 1400 pages long) are published by the European Telecommunications Standards Institute (ETSI).[2]

DVB-MHP adopts the Java programming language and has created the DVB-Java version for broadcast application. Existing application formats can be supported using optional plug – in decoders. [3]

### 2.1. What will MHP be used for?

The MHP specification envisages three modes of use

- Enhanced Broadcasting Profile (One way services)
- Interactive Broadcasting Profile (two-way services)
- Internet Access profile (Internet Services)

This will lead to the availability of such services as:

- Electronic Program Guides (EPG's)
- Super Teletext
- Video Synchronized content
- E-commerce
- Control of Access to TV content

### 2.2. How will MHP be used?

MHP will transport its data through a broadcast stream, which will provide the bulk of the content using broadcast channel protocols. These are defined in DVB standards (EN 301 192 1.2.1) and are transmitted in the same way as conventional TV.

The Interaction Channel will use the Internet Protocol Transmission (IP), which requires a dial-up, wireless or cable TV connection. It is intended that this can in future also be used for purely Internet based service so the Internet is fully accessible through the television.

### 2.3. MHP-compliant Set Top Box (STB)

The MHP places additional requirements upon the standard STB. Figure 1[4] below illustrates the main hardware components of a typical digital TV STB together with the additional components (shaded in blue) needed to support MHP functionality

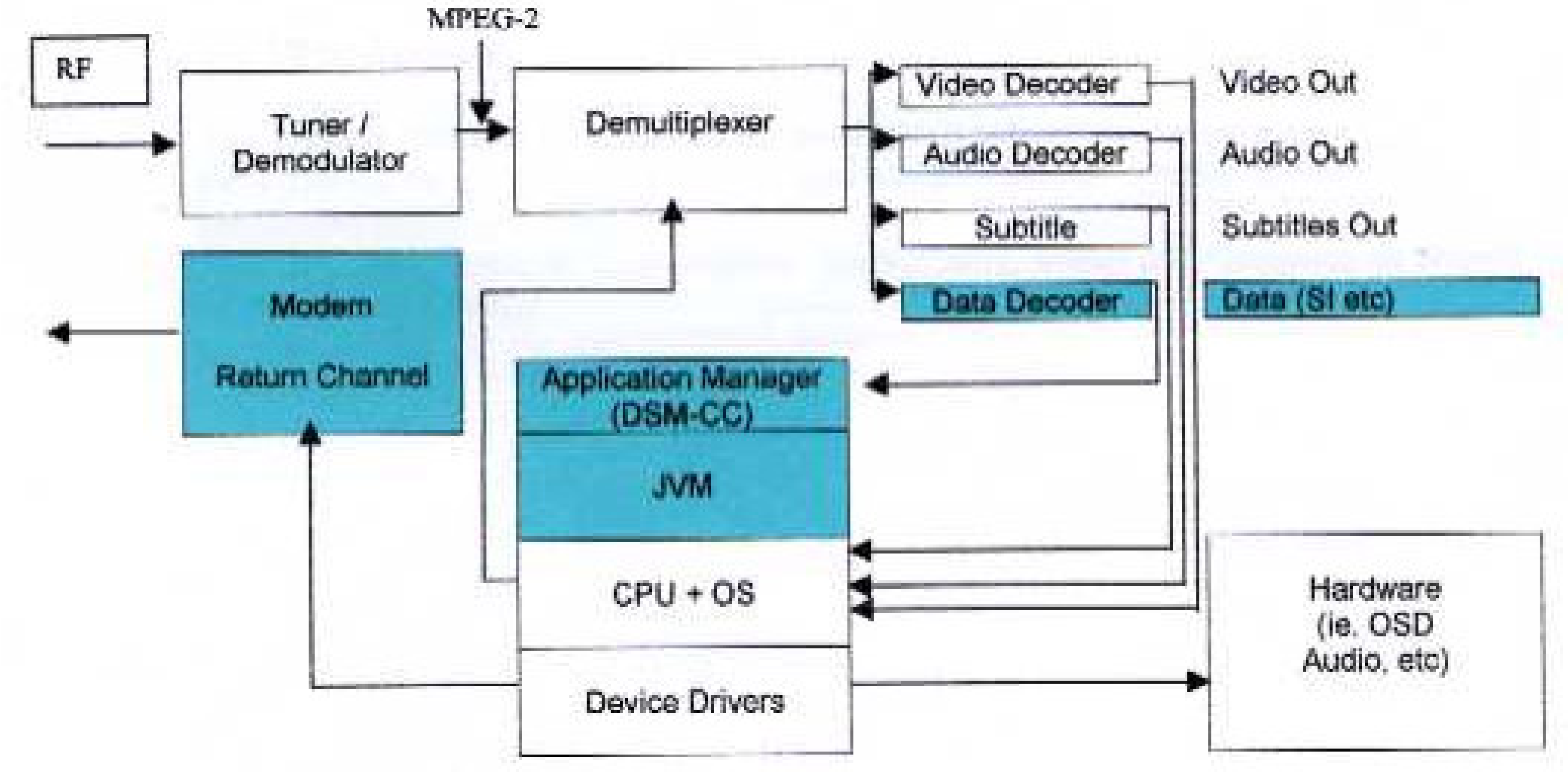

**Figure 1 MHP Compliant STB[4]**

The Radio Frequency (RF) Input (from satellite dish/aerial) is demodulated into an MPEG-2 stream. This is spliced into its video, audio, subtitle, and data streams, and passed on to the CPU and Operating system. The data stream (usually image files) has its own dedicated rendering engine and is directly passed to the on screen display. The data streams are examined for system information, teletext or MHP data object, and then routed via device drivers to the hardware/software modules for interpretation and execution.

### 2.4. What will MHP carry?

Static formats will allow the use of basic images such as JPEG, Gif etc., as well as simple text, MPEG 1 Audio and MPEG 2 Video. Streaming formats involve MPEG Audio, MPEG Video and Subtitles. There is one built in font, which is sans serif and called Tiresias available in four sizes. Downloadable fonts are also available.

MHP will allow the use of Hypertext or Super Teletext referred to as DVB-Html, which is in reality XHTML. Cascading Style sheets are supported, as is a Java script based language called ECMA script. The document object model DOM level 2 is included for the content and structure of documents.
Cookies are supported, and security is available within a sandbox, which limits access to system resources.

### 2.5. MHP Services

The various types of MHP content presented together are called a service. Services contain

Audio, Video, Data and stored material. These can be presented simultaneously.

MHP presents these on three planes: Background, Video and Graphics as shown in figure 2 here[5].

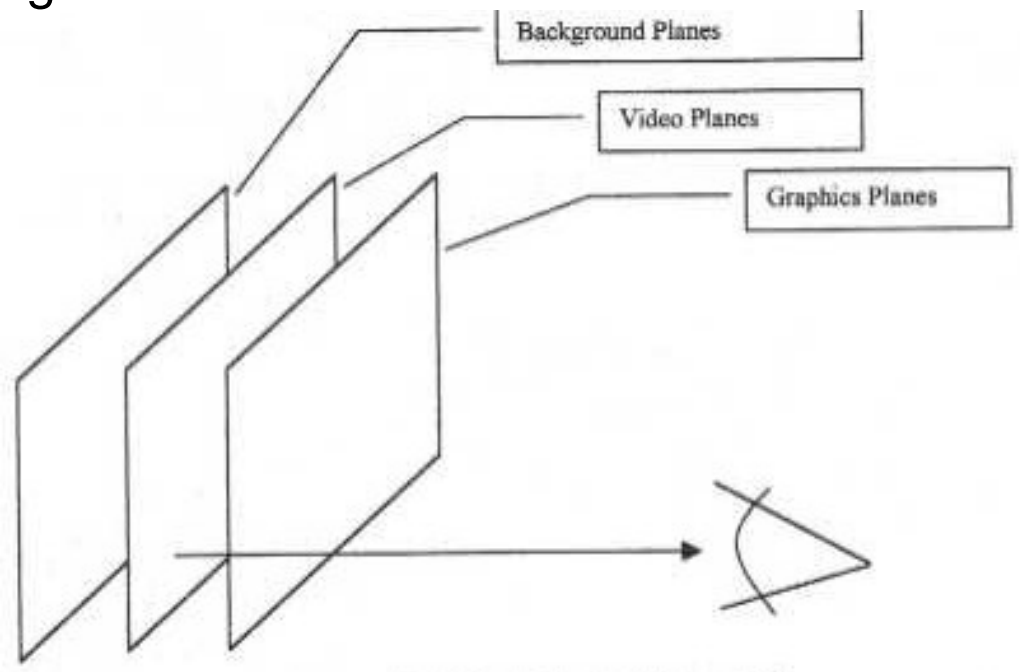

Fig 2: 3-Plane Representation model[6]

Applications are provided with regions of the Graphics plane to place video and graphics. Video is placed on the Video Plane and still images on the Background Plane. Different applications can operate in different regions of the screen in such a way that Video can be displayed on one part of the screen with text appearing in a different region of the screen.

## 3. MHP Architecture

The architecture is in three layers, and at its basic level MHP is set in context as shown in figure 3 below [6]

The MHP receives viewer input via keypad controls. Output communication is presented via a screen and loudspeakers. Data streams are accessible to the MHP, which can write the data to storage (sink or store). The MHP also may have access to remote entities.

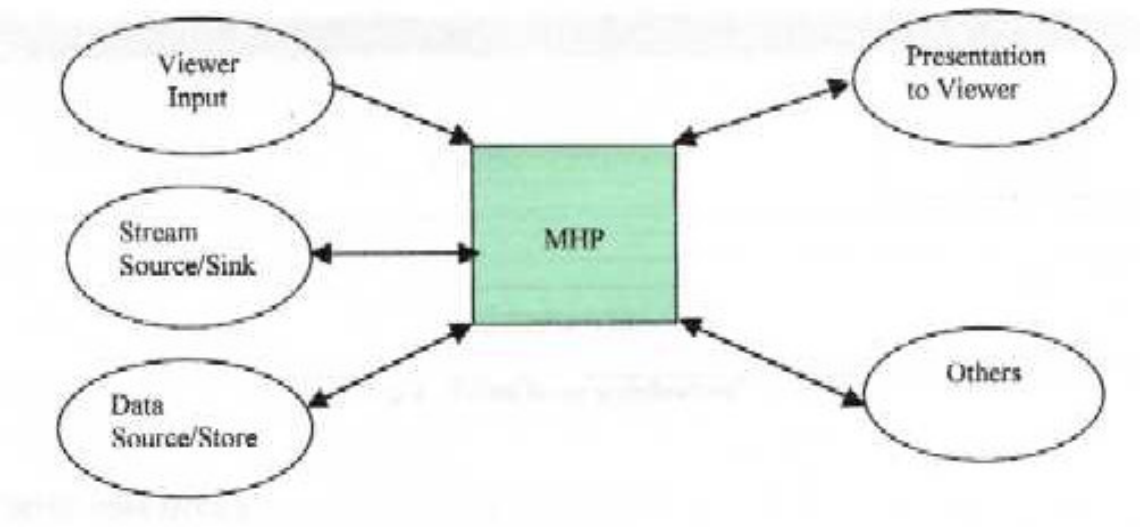

The diagram in figure 4 below shows the software architecture with the various layers.[7]

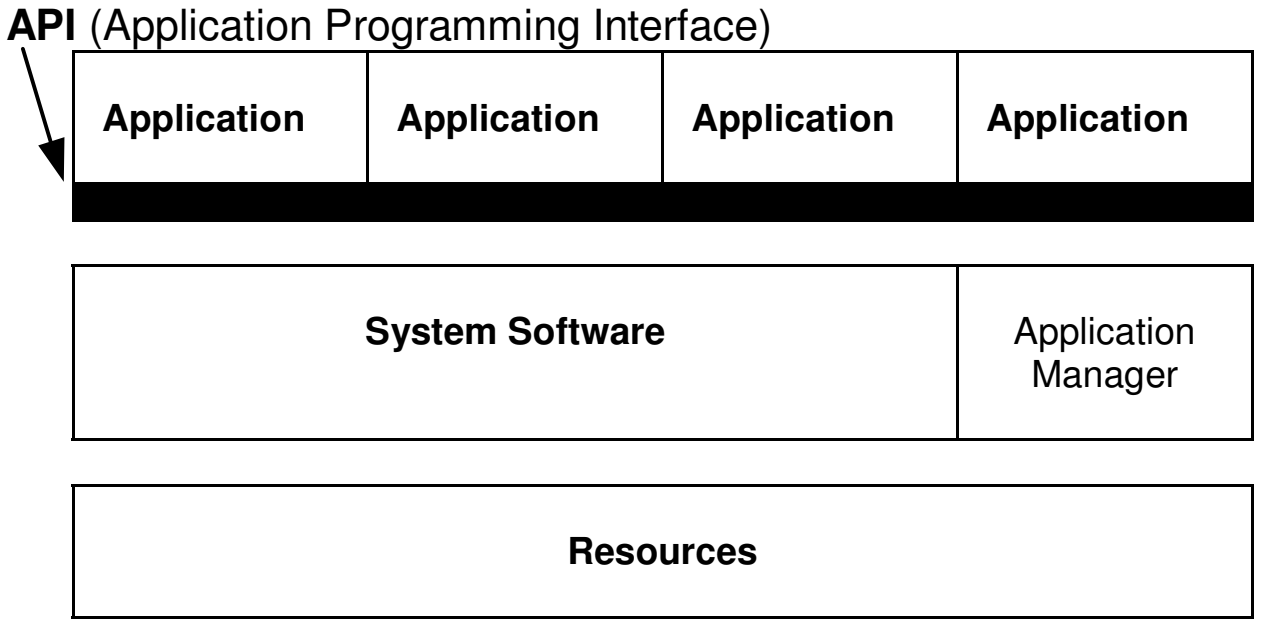

**Figure 4: Three layer architecture**

### 3.1.MHP resources

These include Mpeg processing, I/O devices, CPU, Memory, Graphic systems etc. The resources are provided to the MHP transparently. The application should be able to access all the local resources as though they were part of a single entity.

The following table[8] shows typical hardware requirements of MHP profiles.

| Platform | Processor | RAM | FLASH/ROM |
|---|---|---|---|
| Basic STB | 30 MHz+ | 1-2MB | 1-2 MB |
| MHP Enhanced broadcast profile | 8—130 MHz+ | 8-16 MB | 4 MB |
| MHP Interactive broadcast profile | 80-130 MHz+ | 8-16 MB | 8 MB |
| MHP Internet Access Profile | 150-200 MHz+ | 16-32 MB | 16 MB |

### 3.2.System Software

This includes the real-time Operating System, Java virtual machine and device drivers. Applications are not allowed to address the resources directly. The System software is a middle layer, isolating the application from the Hardware. The purpose of this is to provide portability of the Application so that it can be used with any resources. This enables the available resources to provide an abstract view of the platform to the applications and includes the Application manager.

**3.2.1 Application Manager** (Navigator) The Application Manager runs the life cycle of all MHP compliant applications based on the application signaling information that is embedded in the MPEG-2 transport stream received from the DVB broadcaster.[9]

### 3.3 Applications

Interactive services are implemented by the Applications. These include

- Electronic Program Guides
- Information services
- Play along Games
- E-Commerce
- Secure transactions
- Educational Applications

## 4. MHP System Core

MHP is based around a platform known as DVB- J which includes a Java virtual machine specified by Sun Microsystems. Software Packages provide Application program Interfaces (API's) and MHP gains access to the platform only through the API's.

Applications use the API to access the actual resources of the receiver including databases, streamed media decoders, static content decoders and communications. These resources are functional entities of the receiver and may be finally mapped onto the hardware of the receiver.

### 4.1 Java Virtual Machine (JVM)

With the view of possible future convergence of broadcasting and the Internet Java emerged. It was originally designed for networked environments, has built in support for Internet protocols, and many security features to protect platforms from hostile or damaging applications. Furthermore security is aided by the fact that Java does not permit any direct memory access. The source code for Java applications is compiled to produce Java byte code, which can be interpreted by the Java Virtual machine. Java was attractive to MHP developers because it too does not favour one manufacturer's operating system technology over another. Moreover most competing technologies already used or opted for Java for future technological developments. DVB therefore took the decision to adopt Java for the MHP and developed a version called DVB-J that would be suitable for interactive TV applications.

To sum up, Java has been adopted as the programming language for the MHP for four main reasons:

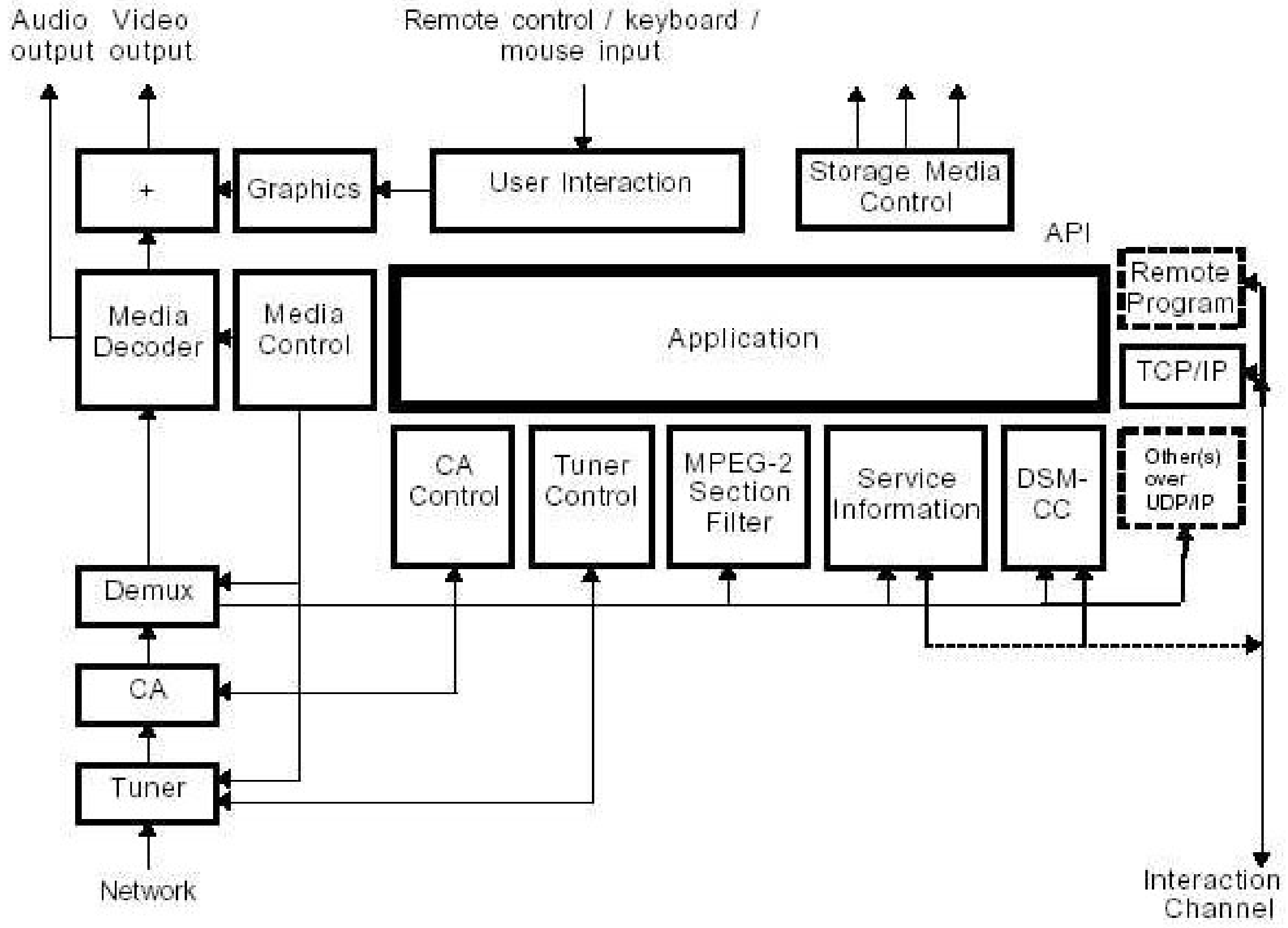

Figure 5 indicates a functional diagram of the enhanced broadcast profile.[10,11]

**i)** It was designed for a networked environment with build in support for TCP/IP.

**ii)** It does not favour one microprocessor over another nor one Operating System over another.

**iii)**. Already competitively and widely developed, which means that interactive TV can benefit from the tools and expertise developed in other industries.[12]

**iv)** Java has a number of built in security features.

**4.2 Transport protocols (TCP/IP)**
MHP makes use of the standard Internet Protocol (IP) for addressing and TCP for transport (see fig.6).

**4.3 Application Programming Interfaces (API's)** (The System Software implements the API's by presenting an abstract model of streams, commands and events, data records, and the hardware resources) The API Layer is the interface between System Software and Applications, and enables interoperability of applications across various MHP implementations provided they conform to DVB-MHP standard. The HTML/Java API defines mechanisms for downloading applications (Xlets) that are embedded in digital broadcast data streams such as MPEG-2 or accessed over Internet Protocol client/server connection. Most MHP compliant receivers include writeable non-volatile storage allowing Xlets to be dynamically upgraded, replaced, or augmented from a remote location.[13] Tools for conditional access (CA), including the DVB common scrambling algorithm for the encryption of transport streams like pay per view programs are available.

**5. MHP Profiles**
MHP provides a consistent set of features in three areas.

**5.1 Enhanced broadcasting one way services.**
This combines digital broadcast of audio/video services with downloaded applications that can enable local interactivity. Enhanced receivers support local interaction including input from the remote control, on-screen-graphical elements and selection from multiple audio/video streams. Enhanced broadcast receivers receive data from a head-end or server possibly carried via a broadcast file system.[14]

MHP allows off-air Xlets to execute by providing them with direct secure limited access to system resources. Xlets can also continue executing during channel changes and query system resources like the hard disks, change channels and pipe data to other media e.g. VCR, DVD or hard disk.

**5.2 Interactive Broadcasting- two way services**
Interactive Broadcast receivers include a return channel to the broadcaster that provides communication with a head-end or server. User interaction is by way of remote control/keyboard. Such receivers are said to be capable of providing electronic commerce, video on demand, email and local chat style communication, and also include capabilities found on enhanced broadcast receivers.[15]

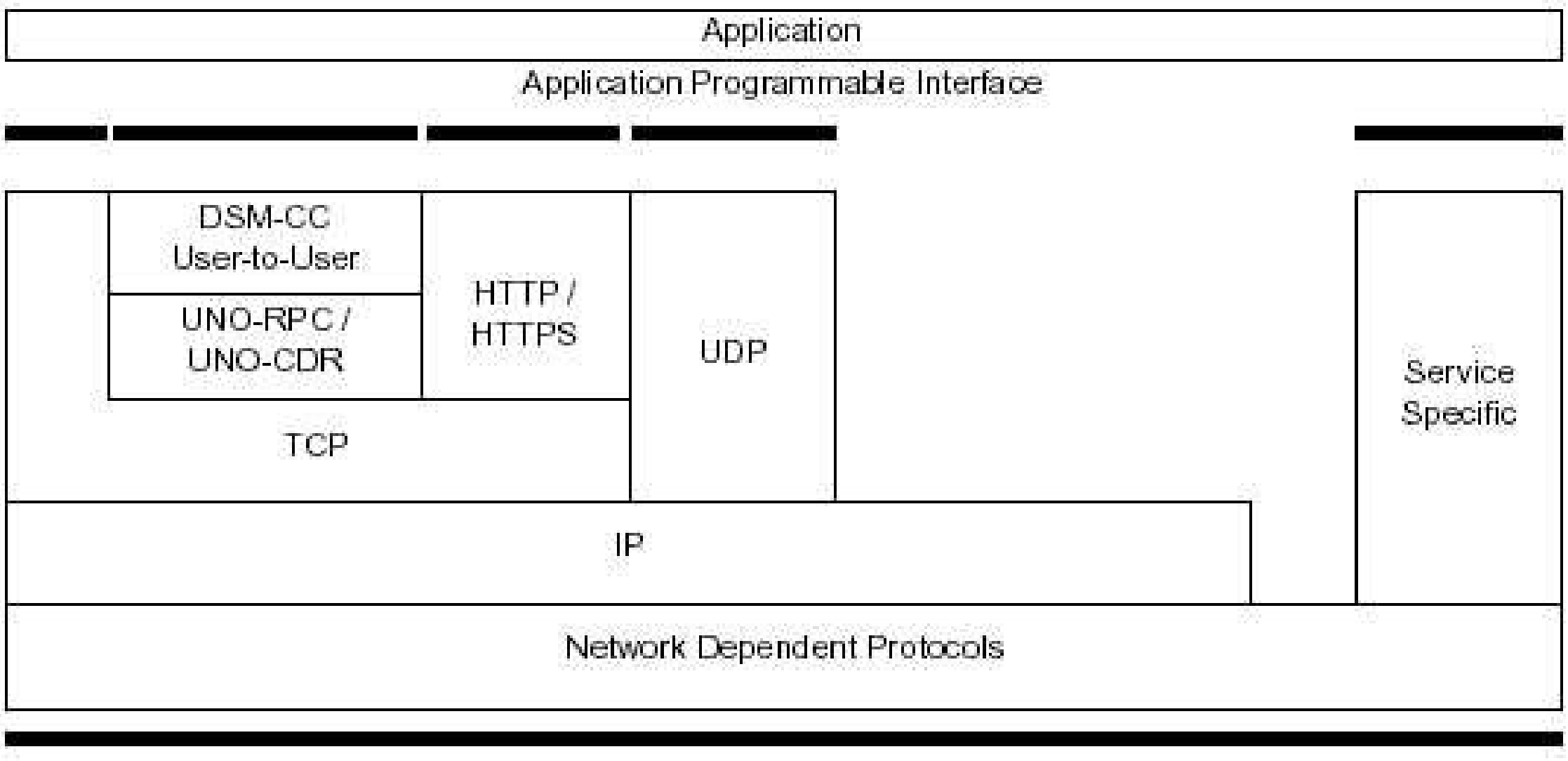

This profile allows user interactivity via an IP return path which enables applications such as instant response advertising e.g. electronic voting, interactive game shows etc.

### 5.3 Internet Access

Internet Access Receivers provide for Internet access and include the capabilities of both enhanced and interactive broadcast receivers. It allows Internet browsing and email capability. Most web pages are designed for minimum resolution of 800x600pixels. Does this profile have a future PC/TV convergence in mind? [16]

However, the display models of a computer and a TV are very different. Integrating the requirements of designing content for the Internet and for a TV will be one of the biggest challenges. [17]

### 5.4 Protocols

The MHP has to communicate through different network types in order to be able to talk to the external world. In order to provided these services the MHP needs to operate various Protocols and support for IP. Broadcast only and interactive services as well as Internet access require the operation of protocol layers as shown in figure 6 above.[18]

The following protocols are required:

- Network dependent protocols which underpin the structure these are defined in ETS 300 800 (CATV), ETS 300 801 (PSTN/ISDN), EN 301 193 (DECT), EN301 195 (GSM), EN 301 199 (LMDS), TR 101 201 (SMATV)

- Internet Protocol (IP)
- Transmission Control Protocol (TCP)
- UNO-RPC (Internet Inter-ORB Protocol (IIOP)
- UNO-CDR (Recordable CD's)
- DCM-CC User to User (defined in ISO/IEC 13818-6

## 6. Summary

MHP is attractive because it is open. For Developers it means ease of implementation. The Platform provides write once, read many operations, as applications will work on any MHP-compliant Set Top Box (STB). For users it means a free choice of STB's from any manufacturer as long as it is MHP compliant.

German Broadcasters (RTL) agreed to rapid introduction of MHP. RTL Newsmedia have developed interactive games to complement its television content with 'Who wants to be a millionaire', introduced TV based email, SMS application for text messaging, interactive news, sport, film, fashion, weather, etc.[19]

Finland started public transmission in August 2000 – the first in the world to do so. With this huge momentum another TV revolution is already upon us. Is this an attempt to turn the TV into a PC? Perhaps not yet.
MHP cares about defining an open middleware standard for interactive TV and attempts to build on top of its existing standards were appropriate. MHP does not appear to be seriously interested to merge European DVB standards with those of the Internet as yet.[20]

[13] Anthony Daniels, Zarlink Semiconductor, http://www.commsdesign.com/design_corner/OEG20020918S0011

[14] Multimedia Home Platform, www.wipro.com/shortcuts/downloads.htm

[15] Multimedia Home Platform, www.wipro.com/shortcuts/downloads.htm

[16] Anthony Daniels, Zarlink Semiconductor, http://www.commsdesign.com/design_corner/OEG20020918S0011

[17] Steven Morris, The Interactive TV Web, http://.www.mhp-interactive.org

[18] MHP Specifications ETSI TS 102 812 V1.1.1 (2001-11), www.etsi.org or http://www.mhp.org/

[19] http://homeplatform.broadcastengineering.com

[20] Steven Morris, The Interactive TV Web http://,www.mhp-interactive.org